# Presolar Silicon Carbide Grains of Types Y and Z: Their Molybdenum Isotopic Compositions and Stellar Origins

Nan Liu[1,2,3*], Thomas Stephan[4,5], Sergio Cristallo[6,7], Roberto Gallino[8], Patrick Boehnke[4,5], Larry R. Nittler[3], Conel M. O'D. Alexander[3], Andrew M. Davis[4,5,9], Reto Trappitsch[4,5,10], Michael J. Pellin[4,5,11], Iris Dillmann[12,13]

[1]Laboratory for Space Sciences and Physics Department, Washington University in St. Louis, St. Louis, MO 63130, USA; nliu@physics.wustl.edu

[2]McDonnell Center for the Space Sciences, Washington University in St. Louis, St. Louis, MO 63130, USA

[3]Department of Terrestrial Magnetism, Carnegie Institution for Science, Washington, DC 20015, USA

[4]Department of the Geophysical Sciences, The University of Chicago, Chicago, IL 60637, USA

[5]Chicago Center for Cosmochemistry, Chicago, IL, USA

[6]INAF-Osservatorio Astronomico d'Abruzzo, Teramo 64100, Italy

[7]INFN-Sezione di Perugia, Perugia 06123, Italy

[8]Dipartimento di Fisica, Università di Torino, Torino 10125, Italy

[9]The Enrico Fermi Institute, The University of Chicago, Chicago, IL 60637, USA

[10]Nuclear and Chemical Sciences Division, Lawrence Livermore National Laboratory, Livermore, CA 94550, USA

[11]Materials Science Division, Argonne National Laboratory, Argonne, IL 60439, USA

[12]TRIUMF, 4004 Westbrook Mall, Vancouver, British Columbia V6T 2A3, Canada

[13]Department of Physics and Astronomy, University of Victoria, Victoria, British Columbia V8P 5C2, Canada

ABSTRACT

We report Mo isotopic compositions of 37 presolar SiC grains of types Y (19) and Z (18), rare types commonly argued to have formed in lower-than-solar metallicity asymptotic giant branch (AGB) stars. Direct comparison of the Y and Z grain data with data for mainstream grains from AGB stars of close-to-solar metallicity demonstrates that the three types of grains have indistinguishable Mo isotopic compositions. We show that the Mo isotope data can be used to constrain the maximum stellar temperatures ($T_{MAX}$) during thermal pulses in AGB stars. Comparison of FRUITY Torino AGB nucleosynthesis model calculations with the grain data for Mo isotopes points to an origin from low-mass (~1.5–3 $M_\odot$) rather than intermediate-mass (>3–~9 $M_\odot$) AGB stars. Because of the low efficiency of $^{22}$Ne($\alpha$,n)$^{25}$Mg at the low $T_{MAX}$ values attained in low-mass AGB stars, model calculations cannot explain the large $^{30}$Si excesses of Z grains as arising from neutron capture, so these excesses remain a puzzle at the moment.

1. INTRODUCTION

Presolar grains—dust condensates which formed in dying stars prior to solar system formation—were identified in primitive meteorites more than 30 years ago based on their large isotopic deviations from the solar system ratios (Zinner et al. 1987). Despite their tiny sizes (sub-μm to μm), the presence of ancient stellar materials on Earth has allowed detailed characterization of their structural and isotopic compositions using a suite of modern microanalytical techniques whose precisions surpass those of the most up-to-date stellar spectroscopic data (see reviews by Zinner 2014 and Nittler & Ciesla 2016). Among diverse presolar mineral phases, silicon carbide (SiC) has been the most extensively studied. Presolar SiC grains are divided into different groups mainly based on their C, N, and Si isotope ratios. Extensive isotopic analyses of multiple elements demonstrate that more than 90% of the grains, known as mainstream (MS) grains, condensed around low-mass (~1.5−3 $M_\odot$), close-to-solar metallicity stars during the asymptotic giant branch (AGB) phase (see review by Zinner 2014), an advanced evolutionary stage of stars with initial mass between ~1.5 $M_\odot$ and ~9 $M_\odot$, where the slow neutron capture process (s-process) takes place (Cameron 1957; Burbidge et al. 1957). The s-process isotopic signatures of trace elements identified in MS type grains are the most convincing evidence that directly links them to low-mass AGB stars (Nicolussi et al. 1997, 1998; Lugaro et al. 2003; Savina et al. 2004). Presolar SiC grains of other types are much rarer (<10% in total), and the stellar origins for many of these types are less clear (e.g., Amari et al. 2001a, 2001b, 2001c). Type X grains (~1−2%), however, are an exception. They have been shown to come from Type II supernovae based on their large excesses of $^{28}$Si and radiogenic $^{26}$Mg, $^{44}$Ca, and $^{49}$Ti from the decay of short-lived $^{26}$Al, $^{44}$Ti, and $^{49}$V, respectively (e.g., Nittler et al. 1996; Hoppe et al. 2000; Liu et al. 2018a). Type AB (~5%) and putative nova grains (<1%), both of which are characterized by $^{12}$C/$^{13}$C≤10 with the latter exhibiting larger $^{15}$N, $^{30}$Si, and inferred $^{26}$Al excesses (Hoppe et al. 1994; Amari et al. 2001c), have quite ambiguous stellar origin(s), and stars of different types have been proposed as their progenitors (e.g., Amari et al. 2001b, 2001c; Fujiya et al. 2013; Pignatari et al. 2015; Liu et al. 2017a). Because the C, N, and Si isotopic signatures of rare types of grains are not sufficient to pin down their stellar origins, isotope data for other elements are required, but, until recently, this has been limited to Mg-Al and Ti in most cases. The urgent need for isotopic data of additional trace elements for these rare presolar SiC grains has been partially satisfied by a series of recent studies (Liu et al. 2016, 2017b, 2018b), in which new insights into their stellar origins were gained based on the obtained isotope data of additional elements for presolar SiC of type AB and putative nova grains.

In comparison to MS grains, Y and Z grains, which represent a few percent each of the whole presolar SiC population, are largely accepted to have also come from AGB stars but with lower metallicities and perhaps with somewhat higher masses (Amari et al. 2001a; Zinner et al. 2006, 2007; Nguyen et al. 2018). Type Y grains have $^{12}$C/$^{13}$C>100 and are more $^{30}$Si-rich than MS grains. Z grains have $^{12}$C/$^{13}$C between 10 and 100, similar to MS grains, and are more $^{30}$Si-rich and $^{29}$Si-poor than Y grains (Alexander 1993; Hoppe et al. 1994, 1997). Regarding $^{14}$N/$^{15}$N and inferred $^{26}$Al/$^{27}$Al ratios, MS, Y, and Z grains are indistinguishable from each other (e.g., Zinner et al. 2006, 2007). The conclusion that Y grains came from AGB stars of roughly 1/2 $Z_\odot$ relied almost solely upon comparison with AGB model predictions to explain their coupled high $^{12}$C/$^{13}$C ratios and large $^{30}$Si excesses (Amari et al. 2001; Zinner et al. 2006). A lower-than-solar metallicity origin for Z grains (~1/3 $Z_\odot$, Zinner et al. 2006) was supported by their subsolar $^{29}$Si/$^{28}$Si and $^{46,47,49}$Ti/$^{48}$Ti ratios, as these isotope ratios are expected to increase during Galactic

Chemical Evolution (GCE) and not be substantially modified by nucleosynthesis in low-mass AGB stars. However, Zinner et al. (2007) were not able to explain the correlated $^{50}$Ti and $^{30}$Si excesses observed in Y and Z grains by neutron capture in low-metallicity AGB models and also noted that the *s*-process efficiency inferred from the Ti isotope data of Y and Z grains seems quite low relative to spectroscopic observations of low-mass stars at ~1/3−1/2 $Z_\odot$ metallicities (Busso et al. 2001).

With the aim of gaining a better understanding of the isotopic signatures of Y and Z grains, we used resonance ionization mass spectrometry to measure Mo isotopes in a large number of Y and Z grains, with the Chicago Instrument for Laser Ionization (CHILI; Stephan et al. 2016). Molybdenum has seven stable isotopes, of which $^{92}$Mo is a pure *p*-process isotope that is bypassed along the *s*-process path, $^{94}$Mo is an almost-pure *p*-process isotope with a marginal *s*-process contribution, and $^{100}$Mo is an almost-pure *r*-process isotope with a marginal *s*-process contribution via $^{99}$Mo$(n,\gamma)^{100}$Mo. On the other hand, the other Mo isotopes are significantly overproduced by the *s*-process during the AGB phase. In the C-rich envelope of AGB stars, apart from $^{92}$Mo, $^{94}$Mo, and $^{100}$Mo, the contribution from the initial Mo isotope abundances incorporated from the interstellar medium is largely overwhelmed by the *s*-process. It is worth pointing out that there exist large uncertainties in the inferred metallicities and masses of the parent stars of Y and Z grains in the literature (Amari et al. 2001; Zinner et al. 2006, 2007), because AGB model predictions for C, Si, Al, and Ti isotope ratios depend strongly on the initial stellar composition, which is largely unknown. In contrast, the Mo isotopic compositions of the Y and Z grains from this study, for the first time, allowed investigation of the *s*-process efficiency in their parent stars without any dependence on the initial stellar composition and thus enable more insight into their stellar origins.

## 2. *s*-PROCESS NUCLEOSYNTHESIS AND UPDATED TORINO MODELS

The reader is referred to Käppeler et al. (2011) and Liu et al. (2018c) for detailed descriptions of *s*-process nucleosynthesis in AGB stars and associated modeling uncertainties. In brief, a thermally pulsing AGB star consists of a C-O core, a He-burning shell, a He intershell, a H-burning shell, and a large convective envelope. The energy necessary to sustain the surface luminosity is provided by the H-burning shell, which is recurrently switched off due to a sudden activation of the He-burning shell. In fact, the energy released by the triple-alpha reaction is large enough to trigger a dynamic runaway (thermal pulse, TP), which causes the He intershell to convect and enriches it with $^{12}$C. If the TP is strong enough to switch off the H-shell (which has been pushed to lower temperatures by thermal expansion of the He intershell), the convective envelope may penetrate into the He intershell, bringing the fresh nucleosynthetic products including $^{12}$C and *s*-process nuclides, to the surface (this mixing episode is known as third dredge-up, TDU). With repeated TDU episodes, the surface C/O ratio increases to above unity, at which point SiC starts to condense (Lodders & Fegley 1995).

During a TDU, some protons are mixed below the formal border of the convective envelope, thus forming a $^{13}$C-rich region via $^{12}$C$(p,\gamma)^{13}$N$(\beta^+v)^{13}$C. This layer, known as $^{13}$C-pocket, represents the major neutron source in low mass AGB stars (Gallino et al. 1998). It radiatively releases neutrons via the $^{13}$C$(\alpha,n)^{16}$O reaction between two adjacent TPs (at ~8 keV, which correspond to ~100 MK). Note that the $^{13}$C pocket must contain a limited amount of $^{14}$N nuclei, which otherwise would capture neutrons via the resonant $^{14}$N$(n,p)^{14}$C reaction and reduce the amount of neutrons available for the production of heavy elements via the *s*-process. In the magnetic-buoyancy-induced $^{13}$C-pocket (Trippella et al. 2016, hereafter Trippella pocket) adopted in this study, the amount of $^{14}$N produced in the $^{13}$C-pocket is extremely limited and thus does not affect the *s*-process efficiency. The $^{14}$N left in the H-burning ashes is mixed within the convective shells generated by TPs, leading to the synthesis of $^{22}$Ne (via a double alpha capture). Thus, a second neutron exposure occurs via the $^{22}$Ne($\alpha$,n)$^{25}$Mg reaction, which works at the bottom of the convective He-shell during TPs at a typical thermal energy of ~23 keV (~300 MK). The $^{22}$Ne abundance in TPs initially depends on the CNO content of the star (i.e., it has a secondary behavior). Once, C is mixed within the envelope through TDU episodes, the $^{22}$Ne abundance approaches a primary behavior due to the increased amount of $^{14}$N left by the H-burning in the He-intershell (Gallino et al. 2006). For reference, in the convective envelope the abundances of $^{12}$C and $^{22}$Ne are overproduced by factors of 12 and 6 at the last TP, respectively, with respect to their initial abundances in the 2 $M_\odot$, 0.72 $Z_\odot$ model (10 TPs in total, final C/O=3.9 in FRUITY[1] stellar models) and factors of 50 and 12, respectively, in the 2 $M_\odot$, 0.22 $Z_\odot$ model (12 TPs in total, final C/O=14.9). The $^{22}$Ne($\alpha$,n)$^{25}$Mg reaction provides a neutron exposure with a neutron density up to $10^9-10^{11}$ cm$^{-3}$ (at least a factor of 100 higher than the neutron density released by the $^{13}$C($\alpha$,n)$^{16}$O), but on a very limited timescale (a few years compared to ~2–20×10$^4$ years between TPs; Cristallo et al., 2011). This results in an additional neutron exposure, which is marginal with respect to the one released in the $^{13}$C pocket. However, this short, high neutron density exposure controls the production of nuclides affected by branch points along the *s*-process path (e.g., Käppeler et al. 2011, Bisterzo et al. 2015), and its neutron density depends strongly on the maximum temperature during TPs ($T_{MAX}$), which increases with increasing stellar mass and decreasing stellar metallicity.

In this study, we derive constraints on $T_{MAX}$ in the parent stars of MS, Y, and Z grains by comparing the grain data to model predictions based on applying the Torino postprocessing method (Gallino et al. 1998) to stellar evolution models computed with the FRUITY code (Cristallo et al. 2009, 2011) for AGB stars with a wide range of masses and metallicities. The postprocessing approach was adopted here in order to efficiently test the effects of nuclear reaction rates and stellar model parameters on model predictions for Si and Mo isotopes. Our ultimate goal is to adopt the constraints obtained using the postprocessing approach and run FRUITY stellar evolutionary models coupled to a full nucleosynthesis network (Cristallo et al. 2009) in the future, which is a more self-consistent approach but extremely time-consuming and computationally demanding. The FRUITY Torino models for AGB stars with metallicities between 0.5 $Z_\odot$ and 1.5 $Z_\odot$ have been recently presented by Liu et al. (2018c). In the present study, we extend the metallicity down to 0.07 $Z_\odot$. In brief, the updated nucleosynthesis calculations are based on physical quantities extracted from FRUITY stellar models (Cristallo et al. 2009, 2011, 2015). Compared to the FRANEC stellar models (Straniero et al. 2003) adopted in previous Torino model calculations, the FRUITY stellar models predict higher TDU efficiencies (Cristallo et al. 2009) and higher mass-loss rates (Straniero et al. 2006). As a consequence, the number of TPs is generally smaller, with the net result to yield lower $T_{MAX}$ and accordingly, a less efficient operation of the $^{22}$Ne($\alpha$,n)$^{25}$Mg reaction.

To be consistent with our previous study (Liu et al. 2018c), FRUITY Torino models shown in this study adopted (1) the lower limit of the $^{22}$Ne($\alpha$,n)$^{25}$Mg rate given by Jaeger et al. (2001) and Longland et al. (2012); and (2) the $^{13}$C pocket profile reported by Trippella et al. (2016). We adopted the $^{22}$Ne($\alpha$,n)$^{25}$Mg rate at its lower limit in this study to investigate its minimum effect in low-metallicity models. The effect of the reaction rate at its upper limit (a factor of two higher, Jaeger et al. 2001; Longland et al. 2012) on model predictions for Si and Mo isotopes was also tested and will be discussed. The Trippella pocket was adopted because it consistently explains the Ni, Sr, and Ba isotopic compositions of MS grains from close-to-solar metallicity AGB stars (Liu et al. 2018c). Compared to the standard $^{13}$C-pocket with an exponential $^{13}$C profile usually adopted in previous studies, the Trippella pocket is larger (in its mass extension), and the $^{13}$C distribution within it is more flattened. A detailed comparison of the Trippella pocket and previously adopted $^{13}$C-pockets (Liu et al. 2014a,b, 2015) is given in Fig. 2 of Liu et al. (2018c).

---

[1] FRUITY stands for Full-Network Repository of Updated Isotopic Tables & Yields (Cristallo et al. 2009).

We adopted Maxwellian-averaged neutron capture cross section (MACS) values reported in both KADoNiS v0.3 and KADoNiS v1.0 databases[2] (Dillmann et al. 2006, 2014) for model calculations. Note that MACS is defined as $\sigma^i_{MACS}=(<\sigma^i(v)v>/v_T)$, where $\sigma^i(v)$ is the $(n,\gamma)$ cross section of a nuclide $i$ and is a function of the relative neutron velocity $v$ and the mean thermal velocity $v_T$. One of the major updates in the KADoNiS v1.0 database is the renormalization of cross sections to the new recommended values for the $^{197}$Au$(n,\gamma)^{198}$Au cross section, which were calculated as the weighted average of the GELINA measurement of Massimi et al. (2014) and the n_TOF measurements of Massimi et al. (2010) and Lederer et al. (2011), since many measurements were carried out relative to this standard. For details, the reader is referred to the "Comment" section in each KADoNiS dataset, which outlines the changes compared to the previous version. In KADoNiS v0.3 (and earlier versions back to the original paper by Bao et al. 2000), the recommended values for isotopes for which only theoretical data are available were derived by normalization to theoretical NON-SMOKER values (Rauscher & Thielemann 2000) of neighboring isotopes. In contrast to this, in the updated KADoNiS v1.0 version, the recommended MACS values for those isotopes were derived from averaging the calculated MACS from the available recently evaluated cross section libraries (e.g., TENDL-2015[3], ENDF/B-VII.1 (Chadwick et al. 2011), JENDL-4.0 (Shibata et al. 2011), JEFF-3.0A (Sublet et al. 2005), and JEFF-3.2[4]) with an estimated 25%−50% uncertainty, depending on the spread of the predicted MACS. This results in differences in the recommended MACS values for the unstable isotopes of Nb and Mo between the two KADoNiS databases (Table 1). For isotopes for which the neutron capture cross sections were measured recently over the full energy range of astrophysical interest (e.g., for the recent data from n_TOF at CERN, GELINA at IRMM Geel, and LANSCE at Los Alamos National Laboratory), the published MACS values were directly implemented into the database without any further need of renormalization.

This method of averaging MACS values calculated from evaluated libraries was also used for the reevaluation of the stable Mo isotopes in KADoNiS v1.0. The libraries include most of the available experimental data and extend the energy range at higher energies with various statistical model calculations. This can lead to slightly different MACS values in different libraries. For the stable Mo isotopes, the experimental data of Musgrove et al. (1976) that were obtained between 3 and 90 keV with high energy resolution ($\Delta E/E \leq 0.2\%$) using the time-of-flight (TOF) method is included in the libraries and governs the recommended MACS values.

---

[2] KADoNiS stands for Karlsruhe Astrophysical Database of Nucleosynthesis in Stars, and the databases are available online at http://www.kadonis.org/ (v0.3) and https://exp-astro.de/kadonis1.0/ (v1.0).
[3] available online at https://tendl.web.psi.ch/tendl_2015/tendl2015.html.
[4] available online at http://www.oecd-nea.org/dbforms/data/eva/evatapes/jeff_32/.

The systematic uncertainties in the Musgrove et al. (1976) data were estimated to be ±4.2% (Winters & Macklin 1987). Evaluating the Musgrove et al. (1976) data, Musgrove et al. (1978) and Winters & Macklin (1987) reported discrepant MACS values for $^{95}$Mo and $^{97}$Mo. Later on, Koehler et al. (2008) remeasured the $^{95}$Mo neutron-capture cross-sections with ORELA using the TOF method, and their calculated MACS values are in good agreement with those reported by Musgrove et al. (1976). While the Winters & Macklin values for stable Mo isotopes were recommended by KADoNiS v0.3, those recommended by KADoNiS v1.0 are the average from recent evaluations[5] (ENDF/B-VII.1, JENDL-4.0, JEFF-3.2, and TENDL-2015), which are 22% and 12% higher for $^{95}$Mo and $^{97}$Mo, respectively, at 30 keV, and closer to the values reported by Musgrove et al. (1978). For $^{98}$Mo and $^{100}$Mo, independent (partially unpublished) activation measurements are available. They agree perfectly in the case of $^{100}$Mo but result in a 17% lower MACS for $^{98}$Mo at 30 keV than presently recommended. This deviation between the two methods for cross section measurements has been observed for many isotopes in KADoNiS and calls for future detailed investigation.

The recommended MACS values in the two databases for the isotopes of interest in this study are shown in Table 1 for comparison. The greatly reduced $^{30}$Si MACS at 8 keV recommended by KADoNiS v1.0 is based on the experimental data of Guber et al. (2003) and was adopted in both FRUITY and FRANEC Torino model calculations presented in this study. The values are given as $\sigma^i_{code}$ values in Table 1 for the following reason. As a first approximation, MACS values of a nuclide at different stellar temperatures are inversely proportional to $v_T$ because of the general $1/v$ behavior of $\sigma^i$. Because $\sigma^i_{code}$ is defined as $<\sigma^i v>v_{T(30\,keV)}$, which is equal to $\sigma^i_{MACS} v_T/v_{T(30\,keV)}$ (Lugaro et al. 2003), the product of $\sigma^i_{code}$ and $v_T$ is directly proportional to the rate of a given neutron capture reaction. Thus, in Table 1, $\sigma^i_{code}$ varies at different stellar temperatures only if $\sigma^i_{MACS}$ deviates from the $1/v_T$ behavior. Such deviations can result in dependencies of the corresponding isotope ratios on the two neutron sources, $^{13}$C and $^{22}$Ne.

## 3. RESULTS

We identified 19 Y and 18 Z grains in this study by imaging thousands of presolar SiC grains, separated from the primitive Murchison meteorite, for C, N, and Si isotopes with the NanoSIMS 50L ion microprobe at the Carnegie Institution. No N isotope data could be obtained in eight of the 37 grains because of a problem with one of the detectors. The Mo isotopic compositions reported here (Table 2) were obtained in the same analytical session as those of 16 AB1, 12 AB2, and 15 MS grains reported by Liu et al. (2017b, 2018b) using CHILI (Stephan et al. 2016). The details for sample preparation and CHILI analysis are given by Liu et al. (2017b) and Stephan et al. (2019), respectively.

We compared the Mo isotopic ratios of the 37 Y and Z grains from this study with those of the 12 MS grains from Liu et al. (2017b, three contaminated MS grains are not included). Figure 1 clearly shows that Y and Z grains indeed carry s-process Mo isotopic signatures, thus confirming their AGB stellar origins. Interestingly, the s-process Mo isotopic compositions of the Y and Z grains are indistinguishable from those of the MS grains. We further examined the data for the three types of grains to see if there exist subtle differences by fitting linear correlations (Mahon, 1996; Trappitsch et al. 2018). Within 1σ uncertainties, the slopes for the three types of grains agree well with each other with the intercepts all being solar. In comparison to literature data from Barzyk et al. (2007) and Stephan et al. (2019), a higher proportion of our grains have close-to-solar Mo composition, likely resulting from their smaller sizes (~0.9 μm in diameter on average) in comparison to the grains (>1 μm) studied in the literature. As a result, more Mo contamination was likely sampled during our RIMS analyses, because the laser beam used for sputtering material was ~1 μm in size (Stephan et al. 2016), comparable to the sizes of the analyzed grains. It is, however, difficult to unambiguously separate contaminated from uncontaminated grains solely based on their Mo isotopic compositions. This is because grains with less negative Mo isotopic compositions could reflect either contamination and/or the composition of their parent stars at the time of their condensation, e.g., less efficient operation of the *s*-process (D3 case in Figs. 1 & 2). On the other hand, Barzyk et al. (2007) showed that multielement isotopic data can be used to identify contaminated grains. Given that correlated Ba and Mo isotope data were obtained in 11 of the 37 grains (Liu et al. 2019), we therefore adopted the method of Barzyk et al. (2007) and identified one contaminated grain, M3-G692, which is highlighted in Table 1 and excluded in all the figures. We cannot exclude the possibility of Mo contamination for the rest of the grains. Mo contamination, however, would not be able to move the three groups of grains to the same linear trends in Fig. 1 if they were either scattered around the lines or had different trends, and thus does not affect any of the discussions in the following section.

---

[5] The cross sections in the evaluated libraries (e.g. ENDFB-VII.1) are based on the measurements of Musgrove et al. (1976).

## 4. DISCUSSION

### 4.1 Constraints on MACS Values of Mo Isotopes

Lugaro et al. (2003) provided a detailed discussion of *s*-process nucleosynthesis in the context of Mo isotope ratios. In Figs. 1 & 2, we compare the three types of grain data to FRUITY Torino model predictions for 2 $M_\odot$ AGB stars with $Z_\odot$ and 0.72 $Z_\odot$, calculated by adopting the two sets of KADoNiS Mo MACS values. Note that we chose 2 $M_\odot$ AGB models for comparison here, because, as will be shown later, AGB model predictions at 2 $M_\odot$ are consistent with the Mo isotope data at a wide range of metallicities (Table 3). It was shown previously that if $^{22}$Ne($\alpha$,n)$^{25}$Mg is not efficiently activated, AGB model predictions for Mo isotope ratios all fall along nearly straight lines in the 3-isotope plots (Lugaro et al. 2003; Barzyk et al. 2007). In Figs. 1a,e, the slopes are almost unity, because $^{92}$Mo is a *p*-only isotope, with $^{94}$Mo mostly a *p*-isotope with a marginal *s*-process contribution, and $^{100}$Mo is mostly an *r*-process isotope with a marginal *s*-process contribution from the channel $^{99}$Mo(n,γ)$^{100}$Mo. As a result, the corresponding ratios of $^{92}$Mo/$^{96}$Mo, $^{94}$Mo/$^{96}$Mo, and $^{100}$Mo/$^{96}$Mo are dominantly determined by the significantly overproduced amount of $^{96}$Mo. On the other hand, in Figs. 1b−d, the slopes are determined by the adopted Mo MACS values. The linear correlations between *s*-process Mo isotope ratios predicted by AGB models are confirmed by the new CHILI Mo isotope data reported here (Fig. 1) and by Stephan et al. (2019). The much less scattered CHILI data are likely a result of the more controlled measurement condition achieved with CHILI (see discussion in Stephan et al. 2019).

The data-model comparisons in Fig. 1 confirm the previous conclusions (Lugaro et al. 2003; Barzyk et al. 2007) that discrepancies exist for $\delta^{95,97,98}$Mo between the grain data and AGB model predictions calculated with the KADoNiS v0.3 MACS values, i.e., those reported by Winters & Macklin (1987). This conclusion is supported by FRUITY Torino model calculations shown in Fig. 1. Observation of data-model comparisons in Fig. 1 further illustrates that the lowered $^{13}$C mass fraction in the D3 case (in which the $^{13}$C mass fraction is reduced by a factor of three) helps to reduce the different slopes in Fig. 1d defined by the grain data and the model predictions in the original case[6], in line with the model predictions shown by Lugaro et al. (2003). The variations in the predicted slopes with varying $^{13}$C mass fractions in Figs. 1b,c,d result from the non-$1/v_T$ behaviors of the MACS values of the corresponding Mo isotope ratios. As a result, their abundances produced by the *s*-process are not simply inversely correlated with their MACS values (Käppeler et al. 2011) and instead show dependencies on the strengths of the $^{13}$C and $^{22}$Ne sources. The amount of $^{96}$Mo produced in the D3 case, however, is too low to match many of the grains, thus failing to explain all the grain data in a consistent way.

The data-model discrepancies, on the other hand, are reduced when the KADoNiS v1.0 values are adopted (Fig. 2). In addition, Fig. 2 shows that when the KADoNiS v1.0 MACS values are adopted, the impact of the chosen $^{13}$C-pocket on predicted Mo isotope ratios is greatly reduced in the plot of $\delta^{98}$Mo vs $\delta^{92}$Mo. This is in better agreement with the fact that the grain data lie along a linear trend with limited scatter in Fig. 2c. The data and model diverge a bit in the plots of $\delta^{95,98}$Mo vs. $\delta^{92}$Mo and may suggest that the true $^{95}$Mo and $^{98}$Mo MACS values are slightly lower than those in KADoNiS v1.0. By reducing the $^{95}$Mo and $^{98}$Mo MACS values by 10% and 5%, respectively, we can reproduce the linear fits to all the grain data shown in Fig. 2. However, it is noteworthy that the model predictions with the original Trippella pocket bend slightly upward during advanced TPs and that the strengths of both deviations are temperature dependent as further detailed below. As a result, higher MACS values (and thus the original KADoNiS v1.0 values) are needed if the $T_{MAX}$ in the parent stars of these grains is slightly higher than those in the models shown in Fig. 2, and vice versa. Overall, the MACS values recommended by KADoNiS v1.0 yield a reasonably good match to the grain data, and, given the small differences (<10%) in the data-model comparison and the effect of $T_{MAX}$, we will therefore adopt the KADoNiS v1.0 values in the models presented in the following discussion.

Finally, Fig. 3a confirms the previous observation (Lugaro et al. 2003) that SiC grains tend to have higher $^{94}$Mo/$^{96}$Mo at low $^{92}$Mo/$^{96}$Mo compared to the model predictions. In the plot of $\delta^{94}$Mo vs $\delta^{92}$Mo in Fig. 1a, FRUITY Torino model predictions overlap well with the linear trend defined by the grain data. However, isotopic ratios are compressed on a linear scale when approaching zero (i.e., delta values approaching −1000‰), and it is therefore difficult to tell if a perfect match to the grain data is reached by the model predictions. For this reason, we plot the grain data in logarithmic scale in Fig. 3 to better examine the data-model comparisons. While the 2 $M_\odot$ model predictions match the grain data well in Fig. 3b, the corresponding predictions in Fig. 3a lie below the grain data, indicating a slightly lower production of $^{94}$Mo in the models compared to the grain data. One of the major channels feeding $^{94}$Mo is via $^{93}$Nb$(n,\gamma)^{94}$Nb$(\beta^+v)^{94}$Mo, and the beta decay rate of $^{94}$Nb has a strong dependence on stellar temperature: its half-life is reduced from 20,000 years at room temperature to 0.5 years at $10^8$ K and 9 days at $3 \times 10^8$ K (Takahashi & Yokoi 1987). Thus, the small data-model discrepancy (<30‰) in Fig. 3a could result from the simplified treatment of the temperature dependence of the $^{94}$Nb beta decay rate, i.e., insufficient resolution, in the postprocessing approach. We will investigate this problem in detail by comparing the grain data with full FRUITY model calculations in the future by adopting the constrained MACS values for Mo isotopes obtained from this study.

---

[6] *Original* case refers to the original $^{13}$C mass fraction and mass extension in the $^{13}$C-pocket reported by Trippella et al. (2016).

*4.2 Contradictory Si and Mo Isotope Ratios of Z Grains*

The Si isotopic compositions of MS, Y, and Z grains have long been interpreted in the literature as the combined products of *s*-process nucleosynthesis and homogeneous GCE, which is thought to move closely along the linear trend defined by MS grains (illustrated in Fig. 4a). Such a linear GCE trend of Si isotopes is also qualitatively supported by model predictions (Timmes & Clayton 1996; Lewis et al. 2013). Y and Z grains plot to the $^{30}$Si-rich side of the MS grain line, which led to the conclusion that a more efficient *s*-process occurred in the parent AGB stars of Y and Z grains (Zinner et al. 2006). In contrast, the $Z_\odot$ and 0.72 $Z_\odot$ FRUITY Torino models shown in Fig. 2 predict only a few tens of ‰ enrichments in $\delta^{30}$Si, which are too low to account for the large $^{30}$Si excesses observed in many of the Z grains. Note that we adopted the Si MACS values recommended by KADoNiS v1.0, where the MACS of $^{30}$Si is greatly reduced at 8 keV (Table 1) with respect to KADoNiS v0.3 and correspond to enhanced $^{30}$Si production by neutron capture during the AGB phase. Thus, the underproduction of $^{30}$Si predicted by the FRUITY Torino models with respect to the Z grain data cannot be explained by the known uncertainties in $^{30}$Si MACS values. In addition, the neutron capture chain, $^{28}$Si$(n,\gamma)^{29}$Si$(n,\gamma)^{30}$Si, ultimately feeds $^{30}$Si. This explains why in the AGB envelope both the original $^{28}$Si and $^{29}$Si abundances remain almost unchanged while $^{30}$Si increases with increasing TDU episodes, as illustrated by the blue line with an arrow in Fig. 4a.

It was previously concluded that Si isotopes are mostly affected by the short, high flux neutron exposure released from $^{22}$Ne$(\alpha,n)^{25}$Mg during TPs (e.g., Zinner et al. 2006). Since $T_{\text{MAX}}$ increases with increasing stellar mass and decreasing stellar metallicity (Cristallo et al. 2009, 2011, 2015), we further compared the grain data with AGB model calculations for 2 $M_\odot$ and 3 $M_\odot$ stars with metallicities down to 0.43 $Z_\odot$ in Figs. 4 & 5, respectively, which, however, still predict only up to 50‰ enrichments in $\delta^{30}$Si. In addition, Fig. 4 shows that 2 $M_\odot$, 0.43 $Z_\odot$ model predictions remain the same even if the $^{22}$Ne$(\alpha,n)^{25}$Mg reaction rate is increased to its upper limit (Jaeger et al. 2001; Longland et al. 2012). Thus, the uncertainty in the $^{22}$Ne$(\alpha,n)^{25}$Mg reaction rate does not affect the data-model comparisons in Figs. 4 & 5. Zinner et al. (2006) also pointed out that the $^{32}$S$(n,\gamma)^{33}$S$(n,\alpha)^{30}$Si reaction could be an important contributor to the abundance of $^{30}$Si in low-mass AGB stars. In our model calculations, the adopted MACS values for $^{32}$S$(n,\gamma)^{33}$S are those recommended by KADoNiS v1.0. Halperin et al. (1980) measured the $^{32}$S$(n,\gamma)^{33}$S cross section from 25 keV up to 1100 keV neutron energy with ORELA using the TOF method, and the corresponding MACS values are recommended by KADoNiS v0.3 with ~5% uncertainty at 30 keV. By averaging the recent evaluation results based on the Halperin et al. (1980) data from different libraries, the recommended MACS values are about 40% higher in KADoNiS v1.0 and result in slightly lowered production of $^{33}$S and in turn lowered production of $^{30}$Si. The difference (<40%), however, is too small to explain the large $^{30}$Si excesses observed in many of the Z grains with respect to MS and Y grains. On the other hand, we normalized the MACS values for $^{33}$S$(n,\alpha)^{30}$Si obtained with TOF by Wagemans et al. (1987) to the activation measurement at 25 keV by Schatz et al. (1995). The renormalized Wagemans et al. (1987) values were adopted in our model calculations. Recently, Praena et al. (2018) measured the neutron capture cross section for this reaction in the neutron energy range from 10 to 300 keV with the n_TOF facility at CERN, but emphasized that new experimental data from thermal value to 10 keV are necessary to better fit the MACS values at low energies. Figure 4 shows that, by increasing the $^{33}$S$(n,\alpha)^{30}$Si reaction rate by a factor of two in the 2 $M_\odot$, 0.43 $Z_\odot$ model (yellow squares), the final prediction for $\delta^{30}$Si only increases from 35‰ to 54‰, which is still too low to explain the majority of the Z grains. Finally, because the abundances of α nuclides are probably enhanced in low-metallicity stars as a result of the GCE effect, we increased the $^{28}$Si and $^{32}$S abundances by 14% relative to the $^{56}$Fe abundance[7] in the 2 $M_\odot$, 0.43 $Z_\odot$ model (purple squares), whose predicted $^{30}$Si enrichments, however, remain the same as the corresponding case with the solar initial isotopic composition (yellow squares). Thus, the various tests shown in Fig. 4 point out that the large $^{30}$Si excesses observed in many of the Z grains cannot be explained by uncertainties in nuclear reaction rates or increased initial abundances of α nuclides in low-metallicity AGB stars.

On the other hand, the 3 $M_\odot$, 0.43 $Z_\odot$ model in Fig. 5, which predicts the largest $\delta^{30}$Si enrichment, shows deviations from the linear trends for Mo isotopes in Figs. 5b,c. The deviations seen in the 3 $M_\odot$, 0.43 $Z_\odot$ model result from the more intense neutron exposure provided by the more efficient operation of the $^{22}$Ne(α,n)$^{25}$Mg reaction at increased $T_{MAX}$. The decreased $\delta^{97,98}$Mo values predicted by this model, however, are not a result of stronger activation of either the $^{95}$Nb or the $^{95}$Zr branch point in this atomic mass region, because the former is expected to lower $\delta^{95}$Mo and the latter to increase $\delta^{97}$Mo (Lugaro et al. 2003). The effect of the $^{95}$Nb branch point is demonstrated by the 3 $M_\odot$, 0.43 $Z_\odot$ model predictions by adopting a reduced $^{95}$Nb MACS in Figs. 5b,c, which results in increased $\delta^{95}$Mo because of the weakened activation of this branch point. The predicted increase in $\delta^{95}$Mo and decrease in $\delta^{97,98}$Mo during advanced TPs in the 3 $M_\odot$, 0.43 $Z_\odot$ model, therefore, cannot be explained by the branching effects. In addition, Lugaro et al. (2003) showed that the $^{95}$Zr branch point barely affects model predictions for Mo isotope ratios at solar metallicity, which was also confirmed by the FRUITY Torino model calculations at metallicities down to 0.22 $Z_\odot$. Instead, these predicted variations in Mo isotope ratios are likely caused by the deviations of their MACS values from the $1/v_T$ rule from 8 keV to 23 keV (Table 1). Examination of $\sigma^i_{code}$ values in Table 1 shows that, in KADoNiS v1.0, the $\sigma^i_{MACS}$ values of $^{98}$Mo show the largest deviation from the $1/v_T$ rule from 8 keV to 23 keV with respect to other stable Mo isotopes. Thus, the non-$1/v_T$ behavior of the $^{98}$Mo MACS values along with the highest s-process production of $^{98}$Mo among all the Mo isotopes except for $^{96}$Mo, results in the strongest dependence of $\delta^{98}$Mo model predictions on the two neutron sources. This is also consistent with the fact that among all Mo isotope ratios model predictions, those for $\delta^{98}$Mo are most sensitive to both the $^{13}$C (Fig. 1d) and $^{22}$Ne sources. In addition, we further show in Fig. 5 that the 3 $M_\odot$, 0.43 $Z_\odot$ model predictions for $\delta^{98}$Mo in the D3 case remain almost unchanged especially during advanced TPs, thus demonstrating that the predicted large deviations in $\delta^{98}$Mo at low metallicities solely result from the enhanced efficiency of the $^{22}$Ne(α,n)$^{25}$Mg reaction and are independent of uncertainties in the $^{13}$C-pocket.

To further investigate the effect of metallicity on the production of $^{30}$Si in AGB stars, we also compared the grain data with model predictions at 0.07 and 0.22 $Z_\odot$ in Fig. 6. Note that Zinner et al. (2006, 2007) did data-model comparisons at metallicities only down to ~1/3 $Z_\odot$ and did not exclude the possibility of even lower metallicities. Figure 6a clearly shows that the production of $^{30}$Si is significantly increased at 0.07 $Z_\odot$ so that the corresponding model predictions can explain the Si isotopic compositions of all the Y and Z grain data except for the three Z grains in green. The 0.07 $Z_\odot$ models, however, predict large shifts from the linear trends in Figs. 6b,c, thus failing to explain the corresponding Mo isotope data. Note that enhanced $\delta^{97}$Mo and $\delta^{98}$Mo values predicted by the 3 $M_\odot$, 0.07 $Z_\odot$ model during advanced TPs (Fig. 6) result from the activation of the $^{95}$Zr branch point. Examination of Figs. 4−6 clearly illustrates that all the models that agree with the Mo isotope data predict only up to 50‰ enrichments in $\delta^{30}$Si and thus cannot explain the Si isotope ratios of the majority of the Z grains. The inconsistent Si and Mo isotope ratios of Z grains are illustrated by the three Z grains (in green) with the highest $^{30}$Si excesses, which overlap well with the rest of the grains in their Mo isotope ratios in Figs. 4−6. In conclusion, the higher $^{12}$C/$^{13}$C and $^{30}$Si/$^{28}$Si ratios of Y grains with respect to MS grains may be accounted for if Y grains came from AGB stars with slightly higher initial masses and/or slightly lower metallicities, because the FRUITY Torino model calculations predict increases in both ratios with decreasing stellar metallicity and increasing stellar mass. The much larger $^{30}$Si enrichments observed in many of the Z grains in Figs. 4−6 by assuming a homogenous linear GCE trend, however, cannot be explained by neutron capture in low-mass AGB stars according to the FRUITY Torino model calculations.

---

[7] Since the abundances of α nuclides are enhanced in the 2 $M_\odot$, 0.43 $Z_\odot$ model, the metallicity, defined as the total abundance of elements heavier than H and He, consequently increases and is higher than 0.43 $Z_\odot$ in this case.

Could the "inconsistent" Si and Mo isotopic compositions of Z grains result from uncertainties in stellar models? In fact, little is known about the mass loss rate in low-mass stars of lower-than-solar metallicity, which can result in large uncertainties in low-metallicity stellar models. To test the effect of stellar model uncertainties, we compared the grain data with FRANEC Torino model predictions, which were calculated by adopting the same set of nuclear reaction rates and the original Trippella pocket as those shown in Figs. 4−6. Consistent with the study of Zinner et al. (2006), Fig. 7a shows that FRANEC Torino model predictions for AGB stars with metallicities of 1/2 $Z_\odot$ and 1/3 $Z_\odot$ can largely explain the $^{30}$Si excesses of the Y and Z grains, respectively. Figures 7b,c, however, clearly show that these FRANEC Torino models (Zinner et al. 2006) fail to explain the well-correlated Mo isotope ratios observed in the grains, again because their $T_{MAX}$ values are too high. The reduced $T_{MAX}$ values in FRUITY Torino models, on the other hand, result from the introduction of an exponentially decaying profile of convective velocities at the inner base of the convective envelope (Straniero et al., 2006; Cristallo et al., 2009), a new mass-loss law calibrated based on a sample of Galactic giant stars (Straniero et al. 2006), and carbon-enhanced molecular opacities (Cristallo et al., 2007). The combined effect of these updates leads to higher TDU efficiencies and higher mass loss rates, and in turn lowered $T_{MAX}$ values (Zinner et al. 2006).

The FRANEC Torino model calculations shown in Fig. 7 further demonstrate that the uncertainties in both the $^{13}$C pocket and the $^{22}$Ne($\alpha$,n)$^{25}$Mg reaction rate cannot help explain the contradictory Mo and Si isotope ratios of the Z grains. Thus, based on Figs. 4–7, we conclude that the inconsistent Si and Mo isotopic compositions of Z grains are unlikely to result from uncertainties in stellar models and in nuclear reaction rates. This conclusion is further supported by the fact that, when matching the Mo isotope data, both FRUITY Torino and FRANEC Torino (during the first few TPs in the case of 3 $M_\odot$, 0.50 $Z_\odot$) model calculations consistently predict up to 50‰ enrichments in $\delta^{30}$Si, which, combined with the effect of homogeneous GCE, can explain the Si isotope data of the MS and Y grains reasonably well. Given that the Si isotope ratios of the grains reflect the effects of both the s-process and GCE and that the effect of the s-process seems quite limited, one possible explanation of the large $^{30}$Si excesses observed in the Z grains is that the initial compositions of their parent stars were anomalously enriched in $^{30}$Si with respect to the homogenous GCE trend shown in Figs. 4−6. It is, however, unclear how their parent stars could have attained such anomalous initial compositions with large $^{30}$Si excesses and negative $\delta^{29}$Si and $\delta^{46,47}$Ti as well. This scenario needs to be examined in the future by considering the yields of Si and Ti isotopes from different stellar sources and by running detailed GCE calculations.

*4.3 Constraints on $T_{MAX}$ in Parent Stars of Y and Z Grains*

Figures 4−6 show that the plot of $\delta^{98}$Mo vs $\delta^{95}$Mo can be used to constrain $T_{MAX}$ and thus to infer the masses and metallicities of the parent stars of Y and Z grains. Table 3 summarizes the data-model comparison results in the plot of $\delta^{98}$Mo vs $\delta^{95}$Mo (most of the models are shown in Figs. 4−6). The Y and Z grain data disagree with the models with cross-marks in Table 3, because the Y and Z grains do not show any observable shifts from the MS grains in their Mo isotope ratios as predicted by these models (Figs. 5 & 6). Moreover, the model-data discrepancies observed in Figs. 5 & 6 are further supported by the grain-model comparison in Fig. 8. Figure 8 shows that all the excluded models in Table 3 based on the plot of $\delta^{98}$Mo vs $\delta^{95}$Mo in Figs. 4−6 also predict upward shifts from the linear trends defined by the grains during advanced TPs as a result of the stronger activation of the $^{99}$Mo branch point, leading to a marginal *s*-process production of *r*-only $^{100}$Mo via $^{99}$Mo$(n,\gamma)^{100}$Mo. In addition, based on Fig. 8, the 2 $M_\odot$, 0.43 $Z_\odot$ model can be further excluded. Thus, the independent constraints on $T_{MAX}$ derived from the data-model comparisons shown in Figs. 4−6 and Fig. 8 are in good agreement with each other and thus quite robust. It is noteworthy that although $T_{MAX}$ plays a dominant role in determining the deviations of model predictions from the linear trends defined by the grain data in the Mo 3-isotope plots, the effect of $T_{MAX}$ is further complicated by variations in the stellar mass and the TP number at which point grain condensation took place. For instance, assuming that all the grains condensed at the last TP in AGB stars of 1.5 $M_\odot$, we can constrain $T_{MAX}$ to lie below 2.88×10$^8$ K in their parent stars ($T_{MAX}$ values of FRUITY stellar models can be found online at http://fruity.oa-teramo.inaf.it/phys_modelli.pl). However, although $T_{MAX}$ exceeds 2.88×10$^8$ K during the last two TPs in the 2 $M_\odot$, $Z_\odot$ model, the corresponding predictions still explain the Mo isotope data as shown in Fig. 2. Thus, it is not straightforward to simply derive an upper limit for $T_{MAX}$ in the parent stars of the grains. The strong effect of $T_{MAX}$, however, can be used to examine the validity of each model in explaining the Mo isotope data, as shown in this study. Finally, we want to stress here that Mo isotopes need to be considered in future studies for grain-model comparison with AGB models using other codes, e.g., the Monash (Karakas & Lattanzio 2014) and NuGrid (Pignatari et al. 2016; Battino et al., 2016) codes, because they are sensitive to $T_{MAX}$ and need to be used to constrain the maximally allowed stellar mass, which is still a matter of debate (e.g., Lugaro et al. 2018, Liu et al. 2018c).

## 5. CONCLUSIONS

The Mo isotope data of the Y and Z grains from this study, for the first time, allowed us to derive independent constraints on the effects of the *s*-process nucleosynthesis in their parent stars. In contrast to $\delta^{30}$Si that depends strongly on its initial composition, $\delta^{95,97,98}$Mo values during the AGB phase are unaffected by their initial compositions and can be used to probe the efficiency of $^{22}$Ne$(\alpha,n)^{25}$Mg during TPs because of its strong dependence on $T_{MAX}$. The Mo isotope data of the grains require relatively weak neutron exposures from $^{22}$Ne$(\alpha,n)^{25}$Mg during TPs. The inferred low temperatures point to parent AGB stars with relatively low-masses and argue against the intermediate-mass AGB stellar origins of Y grains as previously suggested. Because the $^{30}$Si production is mostly controlled by $^{22}$Ne$(\alpha,n)^{25}$Mg, the large $^{30}$Si excesses observed in Z grains, therefore, cannot be explained by neutron capture in low-mass AGB stars due to the inefficient operation of $^{22}$Ne$(\alpha,n)^{25}$Mg. In our future work, we will continue the investigation by comparing Y and Z grains with AGB models for Sr and Ba isotopes (Liu et al. 2019) to better constrain the stellar metallicities of their parent AGB stars.

Acknowledgements: This work was supported by NASA through grants NNX10AI63G and NNX17AE28G to L.R.N. and NNX15AF78G and 80NSSC17K0251 to A.M.D. Part of this work was performed under the auspices of the U.S. Department of Energy by Lawrence Livermore National Laboratory under Contract DE-AC52-07NA27344. LLNL-JRNL-765107.

## FIGURE CAPTIONS

Fig. 1. Molybdenum 3-isotope plots comparing MS, Y, and Z grains with FRUITY Torino model predictions by adopting the Trippella pocket and KADoNiS v0.3 MACS values. The entire evolution of the AGB envelope composition is shown, but symbols are plotted only when C>O. The numbers in the Trippella pocket labels in parentheses in the legend refer to the extensions of the $^{13}$C-pocket from the bottom of the He envelope to the top of the He-intershell in mass coordinate. *Original* and *D3* cases refer to the original $^{13}$C mass fraction reported by Trippella et al. (2016) and the $^{13}$C mass fraction lowered by a factor of three, respectively. Errors are 2σ.

Fig. 2. The same as Fig. 1 but the model predictions were run with MACS values from KADoNiS v1.0.

Fig. 3. Mo 3-isotope plots in logarithmic scale comparing the same set of grain data in Figs. 1 & 2 to 2 $M_\odot$ model predictions. Errors are 2σ.

Fig. 4. Three-isotope plots of (a) δ($^{29}$Si/$^{28}$Si) vs δ($^{30}$Si/$^{28}$Si), (b) δ($^{97}$Mo/$^{96}$Mo) vs δ($^{95}$Mo/$^{96}$Mo), and (c) δ($^{98}$Mo/$^{96}$Mo) vs δ($^{95}$Mo/$^{96}$Mo) comparing MS, Y, and Z grains with 2 $M_\odot$ FRUITY Torino model predictions. Three Z grains with the largest enrichments in $^{30}$Si are highlighted in green. In panel (a), we included ~300 additional MS grains with high-precision Si isotope data (denoted as MS (all), error bars are not shown for clarity) for comparison with Y and Z grains, all of which were analyzed in the same NanoSIMS sessions. The red and blue lines with arrows are illustration instead of actual model predictions. Errors are 2σ.

Fig. 5. The same set of Mo isotope plots as in Fig. 4 with the grains compared to 3 $M_\odot$ FRUITY Torino model predictions.

Fig. 6. The same set of Mo isotope plots as in Figs. 4 & 5 with the grains compared to FRUITY Torino model predictions at 0.07 and 0.22 $Z_\odot$.

Fig. 7. The same set of plots as in Figs. 4−6 with the grains compared to FRANEC Torino model predictions for 3 $M_\odot$ AGB stars with different metallicities.

Fig. 8. Three-isotope plot of [$^{100}$Mo/$^{96}$Mo] vs [$^{92}$Mo/$^{96}$Mo] with the grains compared to FRUITY Torino predictions shown in Figs. 2−6.

Table 1. Silicon, Niobium, and Molybdenum Neutron Capture Cross Sections, $\sigma^i_{code}$=<$\sigma^i v$>$v_{T(30\ keV)}$=$\sigma^i_{MACS}\ v_T/v_{T(30\ keV)}$ (mbarn)

Table 2. Isotope Data of Types Y and Z Grains

Table 3. Summary of a Complete Model-Data Comparison at Different Metallicities and Masses

## REFERENCES

Alexander, C. M. O'D. 1993, GeCoA, 57, 2869

Amari, S., Nittler, L. R., Zinner, E., et al. 2001a, ApJ, 546, 248

Amari, S., Nittler, L. R., Zinner, E., Lodders, K., & Lewis, R. S. 2001b, ApJ, 559, 463

Amari, S., Gao, X., Nittler, L. R., & Zinner, E. 2001c, ApJ, 551, 1065
Barzyk, J. G., Savina, M. R., Davis, A. M., et al. 2007, M&PS, 42, 1103
Bao, Z. Y., Beer, H., Käppeler F., et al. 2000, ADNDT, 76, 70
Bisterzo, S., Gallino, R., Käppeler, et al. 2015, MNRAS, 449, 506
Burbidge, E. M., Burbidge, G. R., Fowler, W. A., & Hoyle, F. 1957, RvMP, 29, 547
Busso, M., Gallino, R., Lambert, D. L., et al. 2001, ApJ, 557, 802
Cameron, A. G. W. 1957, A.E.C.L. Chalk River, Canada, Technical Report No. CRL-41
Chadwick, M.B., Herman. M., Obložinský, P. et al. 2011, NDS, 112, 2887
Cristallo, S., Straniero, O., Gallino, R., et al. 2009, ApJ, 696, 797
Cristallo, S., Piersanti, L., Straniero, O., et al. 2011, ApJS, 197, 2
Cristallo, S., Straniero, O., Piersanti, L., & Gobrecht, D. 2015, ApJS, 219, 40
Dillmann, I., Heil, M., Käppeler, F., Plag, R., et al. 2006, 2006AIPC 819, Capture Gamma-Ray Spectroscopy and Related Topics, ed. A. Woehr & A. Aprahamian (Melville, NY: AIP), 123
Dillmann, I., Plag, R., Käppeler, F., et al. 2014, 2014nic conf, XIII Nuclei in the Cosmos (NIC XIII), 57
Fujiya W., Hoppe P., Zinner E., et al. 2013, ApJL, 776, L29
Gallino, R., Arlandini, C., Busso, M., et al. 1998, ApJ, 497, 388
Gallino, R., Busso, M., Picchio, G., & Raiteri, C. M. 1990, Natur, 348, 298
Gallino, R., Bisterzo, S., Husti, L., et al. 2006, 2006nic conf, IX Nuclei in the Cosmos (NIC XIII), 100
Halperin, J., Johnson, C. H., Winters, R. R., & Macklin, R. L. 1980, PhRvC, 21, 545
Hoppe, P., Amari, S., Zinner, E., Ireland, T., & Lewis, R. S. 1994, ApJ, 430, 870
Hoppe, P., Annen, P., Strebel, R., et al. 1997, ApJ, 487, L101
Hoppe, P., Strebel, R., Eberhardt, P., Amari, S., & Lewis, R. S. 2000, M&PS, 35, 1157
Jaeger, M., Kunz, R., Mayer, A., et al. 2001, PhRvL, 87, 202501
Käppeler, F., Gallino, R., Bisterzo, S., & Aoki, W. 2011, RvMP, 83, 157
Karakas, A. I., & Lattanzio, J. C. 2014, PASA, 31, e030
Koehler, P. E., Harvey, J. A., Guber, K. H, & Wiarda, D., 2008, 2008nic conf, X Nuclei in the Cosmos (NIC X), 41
Lederer, C., Colonna, N., Domingo-Pardo, C., et al. 2011, PhRvC, 83, 034608
Lewis, K. M., Lugaro, M., Gibson, B. K., & Pilkington, K. 2013. ApJL, 768, L19
Liu, N., Stephan, T., Cristallo, S., 2019, LPI 50, #2132
Liu, N., Nittler, L. R., Alexander, C. M. O'D., & Wang, J. 2018a, SciA, 4, eaao1054
Liu, N., Stephan, T., Boehnke, P., et al. 2018b, ApJ, 855, 144
Liu, N., Gallino, R., Cristallo, S., et al. 2018c, ApJ, 865, 112
Liu, N., Nittler, R. L., Pignatari, M., Alexander, C. M. O'D., & Wang, J. 2017a, ApJL, 842, L1
Liu, N, Stephan, T., Boehnke, P., et al. 2017b, ApJL, 844, L12

Liu, N., Nittler, L. R., Alexander, C. M. O'D., et al. 2016, ApJ, 820, 140
Liu, N., Savina, M. R., Gallino, R., et al. 2015, ApJ, 803, 12
Liu, N., Gallino, R., Bisterzo, S., et al. 2014a, ApJ, 788, 163
Liu, N., Savina, M. R., Davis, A. M., et al. 2014b, ApJ, 786, 66
Lodders, K., & Fegley, B., Jr. 1995, Metic, 30, 661
Longland, R., Iliadis, C., & Karakas, A. I. 2012, PhRvC, 85, 065809
Lugaro, M., Davis, A. M., Gallino, R., et al. 2003, ApJ, 593, 486
Massimi, C., Domingo-Pardo, C., Vannini, G., et al. 2010, PRC, 81, 044616
Massimi, C., Becker, B., Dupont, E., et al. 2014, EPJA, 50, 124
Mahon, K. I. 1996, IGRv, 38, 293
Musgrove, A. de L., Allen, B. J., Boldeman, J. W. & Macklin, R. L. 1976, NucPhA 270, 108
Musgrove, A. de L., Allen, B., and Macklin, R. 1978, in Neutron Physics and Nuclear Data for Reactors and Other Applied Purposes (OECD, Paris, 1978), 426.
Nicolussi, G. K., Davis, A. M., Pellin, M. J., et al. 1997, Sci, 277, 1281
Nicolussi, G. K., Pellin, M. J., Lewis, R. S., et al. 1998, GeCoA, 62, 1093
Nittler, L. R., Amari, S., Zinner, E., Woosley, S. E., & Lewis, R. S. 1996, ApJ, 462, L31
Nittler, L. R., & Ciesla, F. 2016, ARA&A, 54, 53
Nguyen, A. N., Nittler, L. R., Alexander, C. M. O'D., & Hoppe, P. 2018, GeCoA, 221, 162
Pignatari, M., Zinner, E., Hoppe, P., et al. 2015, ApJL, 808, L43
Pignatari, M., Herwig, F., Hirschi, R., et al. 2016, ApJS, 225, 54
Praena, J., Sabaté-Gilarte, M., Porras, I., et al. 2018, PhRvC, 97, 064603
Rauscher, T., & Thielemann, F.-K. 2000, ADNDT, 75, 1
Savina, M. R., Davis, A. M., Tripa, C. E., et al. 2004, Sci, 303, 649
Schatz, H., Jaag, S., Linker, G., Steininger, R., & Käppeler, F. 1995, PhRvC, 51, 379
Shibata, K., Iwamoto, O., Nakagawa, T., et al. 2011, J. Nucl. Sci. Technol., 48, 1
Stephan, T., Trappitsch, R., Davis, A. M., et al. 2016, IJMSp, 407, 1
Stephan, T., Trappitsch, R., Hoppe, P., et al. 2019, ApJ, 877, 101
Straniero, O., Domínguez, I., Cristallo, S., & Gallino, R. 2003, PASA, 20, 389
Straniero, O., Gallino, R., & Cristallo, S. 2006, NucPhA, 777, 311
Sublet, J.-Ch., Koning, A. J., Forrest, R. A., & Kopecky, J. 2005, AIPC 769, International Conference on Nuclear Data for Science and Technology, 203, available online: http://www.oecd-nea.org/dbforms/data/eva/evatapes/jeff_32/
Takahashi, K., & Yokoi, K. 1987, ADNDT, 36, 375
Timmes, F. X., & Clayton, D. D. 1996, ApJ, 472, 723
Trappitsch, R., Boehnke, P., Stephan, T., et al. 2018, ApJ, 857, L15F
Trippella, O., Busso, M., Palmerini, S., Maiorca, E, & Nucci, M. C. 2016, ApJ, 818, 125
Wagemans, C., Weigmann, H., & Barthelemy, R. 1987, NucPh, A469, 497
Winters R., & Macklin, R. L. 1987 ApJ, 313, 808

Woosley, S. E., Fowler W. A., Holmes, J. A., & Zimmerman, B. A. 1978, ADNDT, 22, 371
Zinner, E., Ming, T., & Anders, E. 1987, Natur, 330, 730
Zinner, E., Nittler, L. R., Gallino, R., & Karakas, A. I. 2006, ApJ, 650, 350
Zinner, E., Amari, S., Guinness, R., et al. 2007, GeCoA, 71, 4786
Zinner, E. 2014, in Treatise on Geochemistry, Vol. 1, ed. A. M. Davis, (Elsevier, Oxford), 181

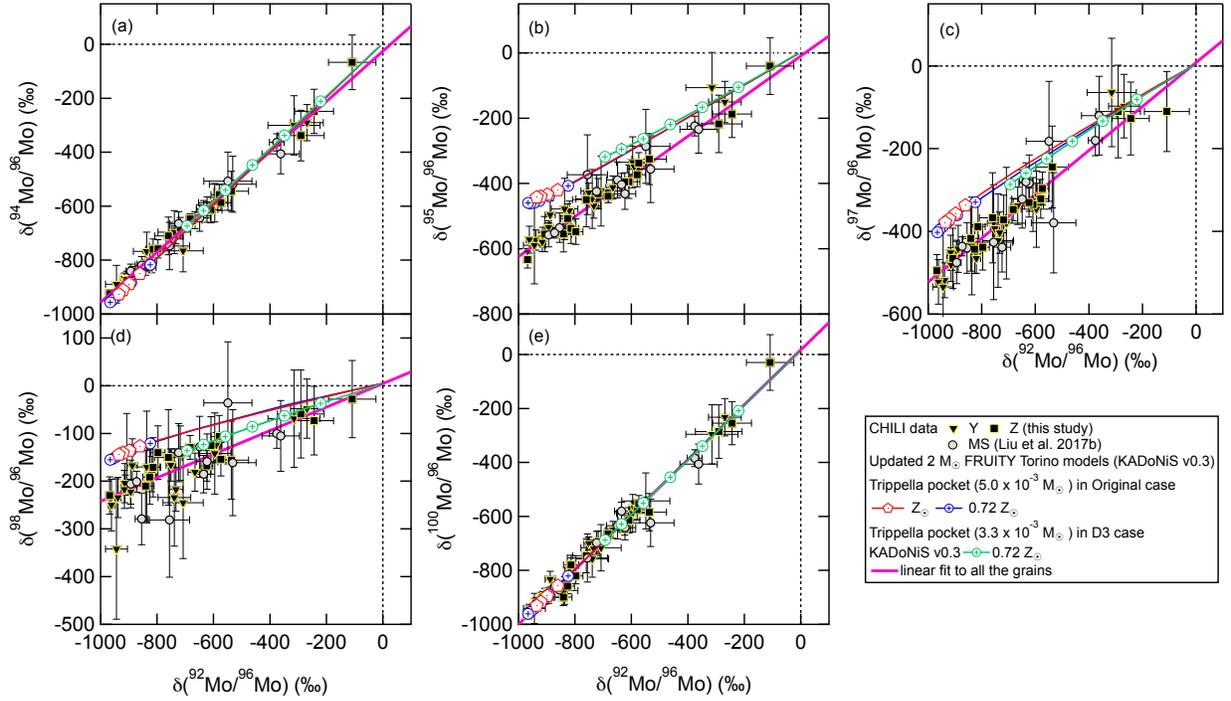

Fig. 1. Molybdenum 3-isotope plots comparing MS, Y, and Z grains with FRUITY Torino model predictions by adopting the Trippella pocket and KADoNiS v0.3 MACS values. The entire evolution of the AGB envelope composition is shown, but symbols are plotted only when C>O. The numbers in the Trippella pocket labels in parentheses in the legend refer to the extensions of the $^{13}$C-pocket from the bottom of the He envelope to the top of the He-intershell in mass coordinate. *Original* and *D3* cases refer to the original $^{13}$C mass fraction reported by Trippella et al. (2016) and the $^{13}$C mass fraction lowered by a factor of three, respectively. Errors are 2σ.

The Astrophysical Journal

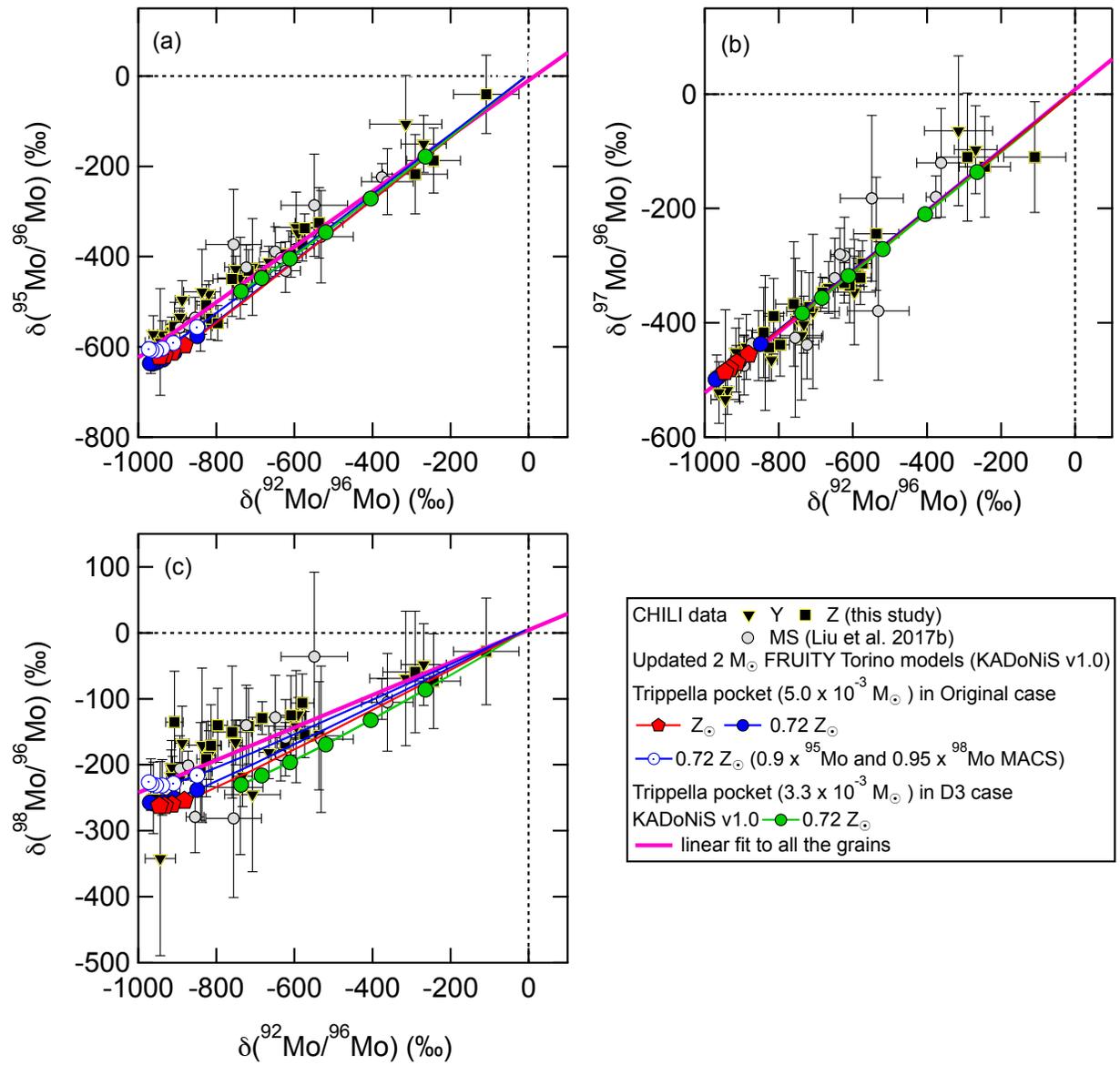

Fig. 2. The same as Fig. 1 but the model predictions were run with MACS values from KADoNiS v1.0.

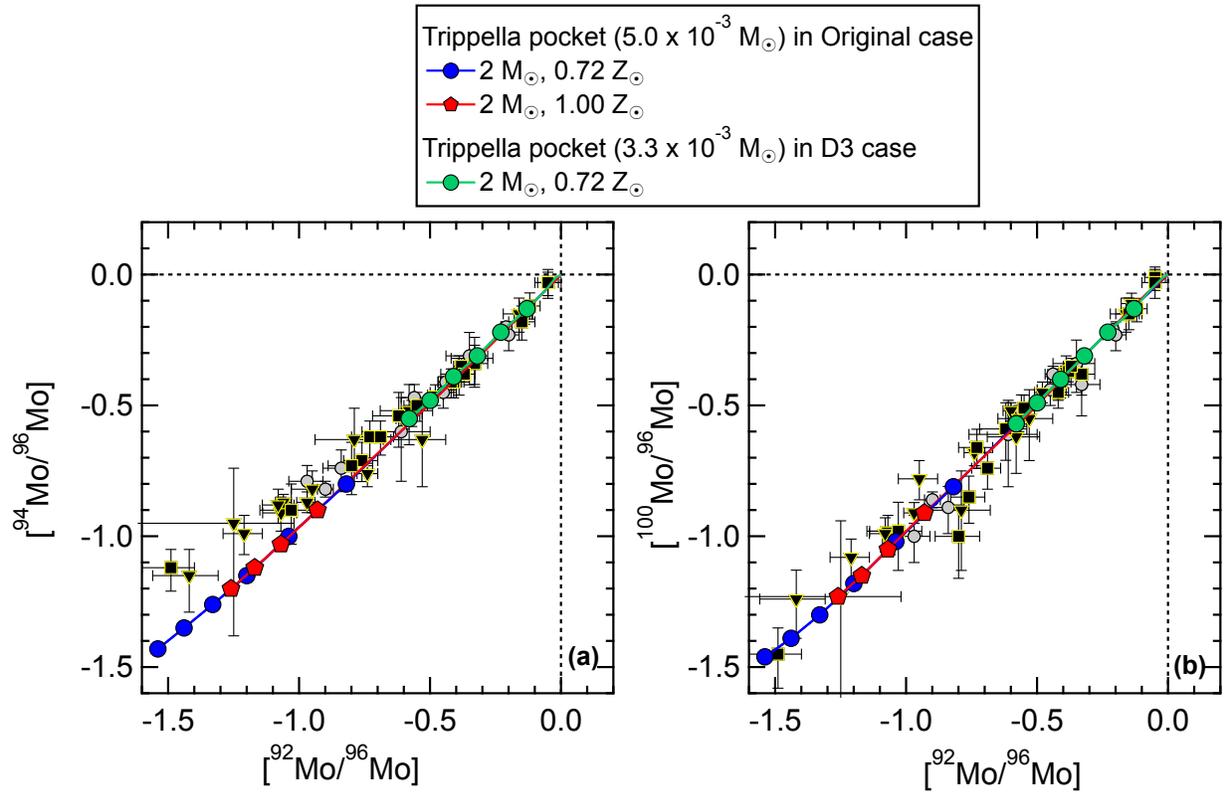

Fig. 3. Mo 3-isotope plots in logarithmic scale comparing the same set of grain data in Figs. 1 & 2 to 2 $M_\odot$ model predictions. Errors are 2σ.

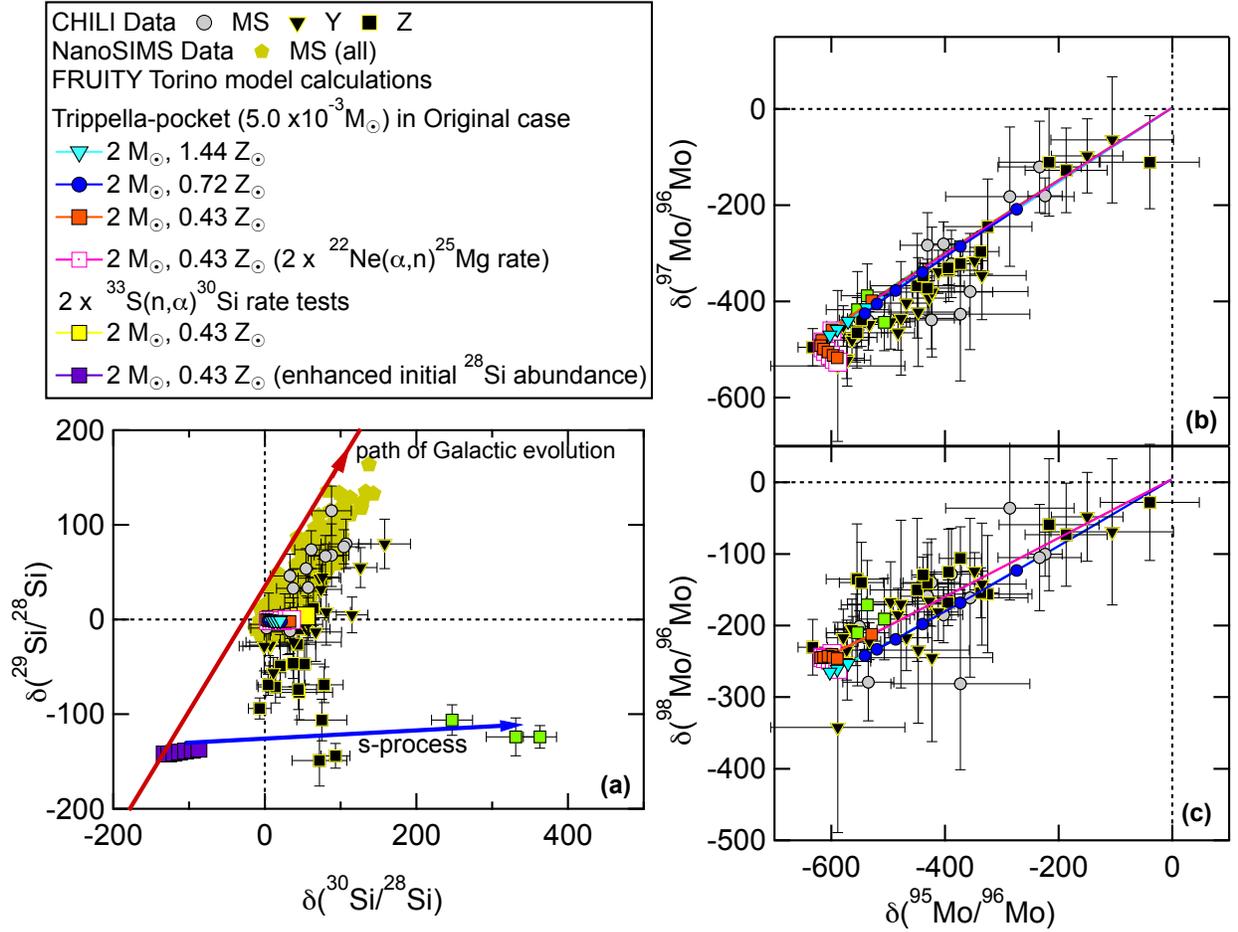

Fig. 4. Three-isotope plots of (a) $\delta(^{29}Si/^{28}Si)$ vs $\delta(^{30}Si/^{28}Si)$, (b) $\delta(^{97}Mo/^{96}Mo)$ vs $\delta(^{95}Mo/^{96}Mo)$, and (c) $\delta(^{98}Mo/^{96}Mo)$ vs $\delta(^{95}Mo/^{96}Mo)$ comparing MS, Y, and Z grains with 2 $M_\odot$ FRUITY Torino model predictions. Three Z grains with the largest enrichments in $^{30}Si$ are highlighted in green. In panel (a), we included ~300 additional MS grains with high-precision Si isotope data (denoted as MS (all), error bars are not shown for clarity) for comparison with Y and Z grains, all of which were analyzed in the same NanoSIMS sessions. The red and blue lines with arrows are illustration instead of actual model predictions. Errors are 2σ.

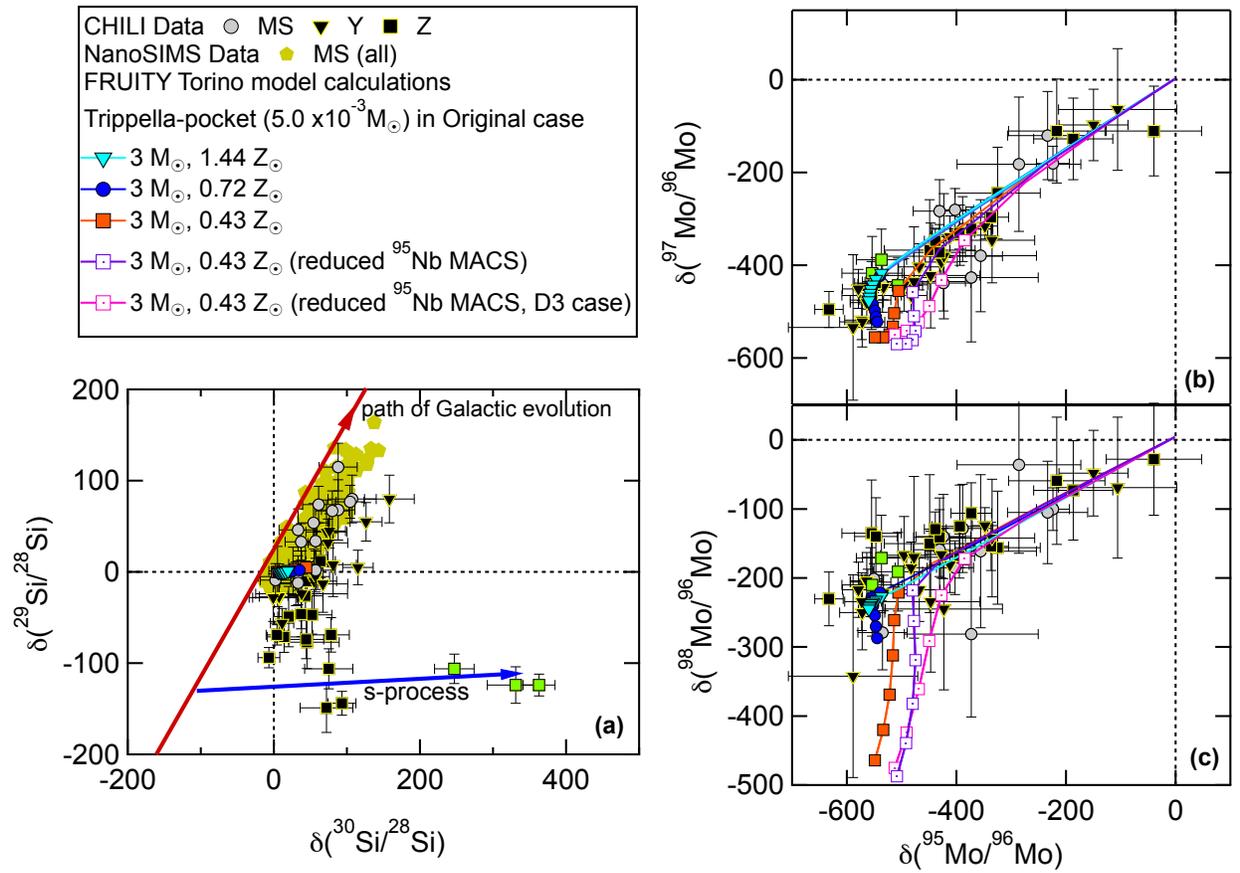

Fig. 5. The same set of Mo isotope plots as in Fig. 4 with the grains compared to 3 $M_\odot$ FRUITY Torino model predictions.

The Astrophysical Journal

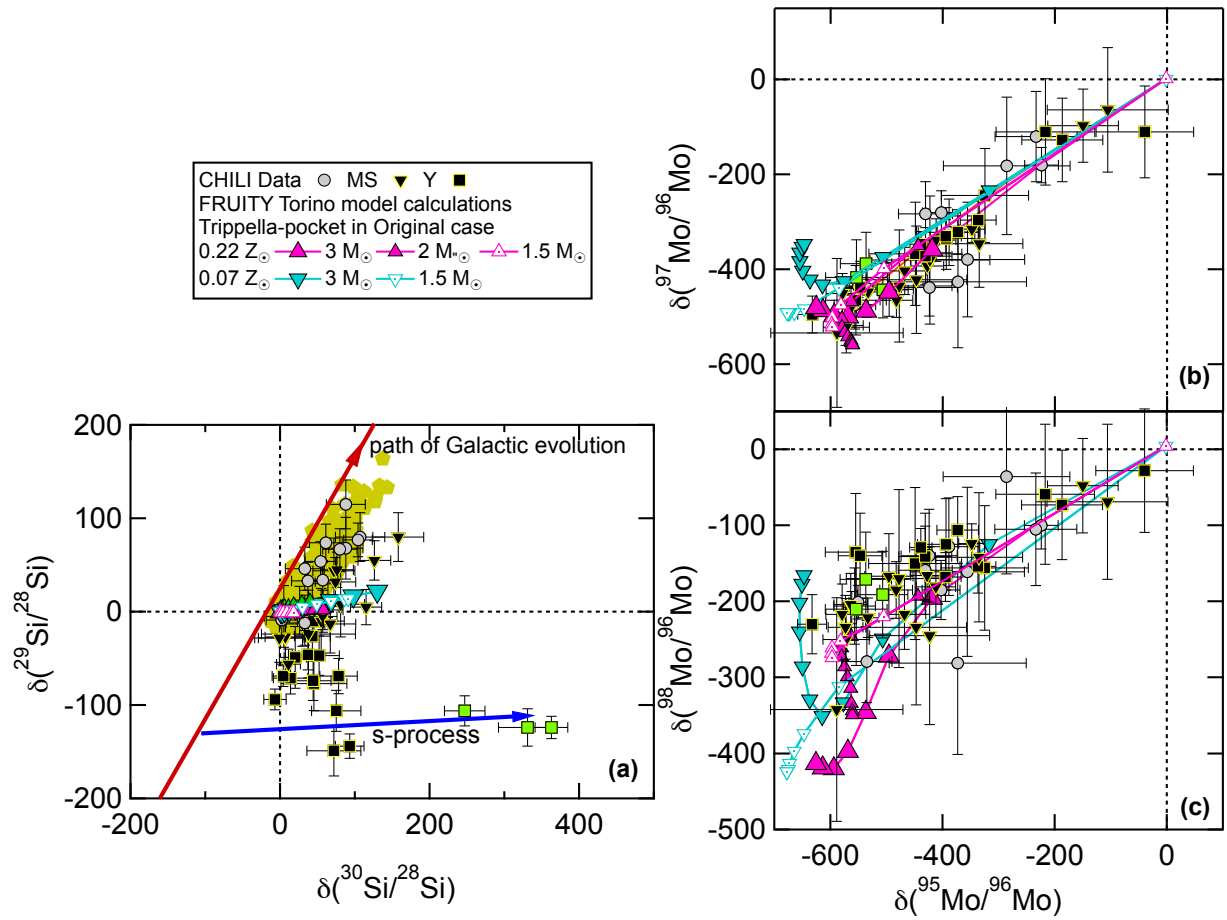

Fig. 6. The same set of Mo isotope plots as in Figs. 4 & 5 with the grains compared to FRUITY Torino model predictions at 0.07 and 0.22 $Z_\odot$.

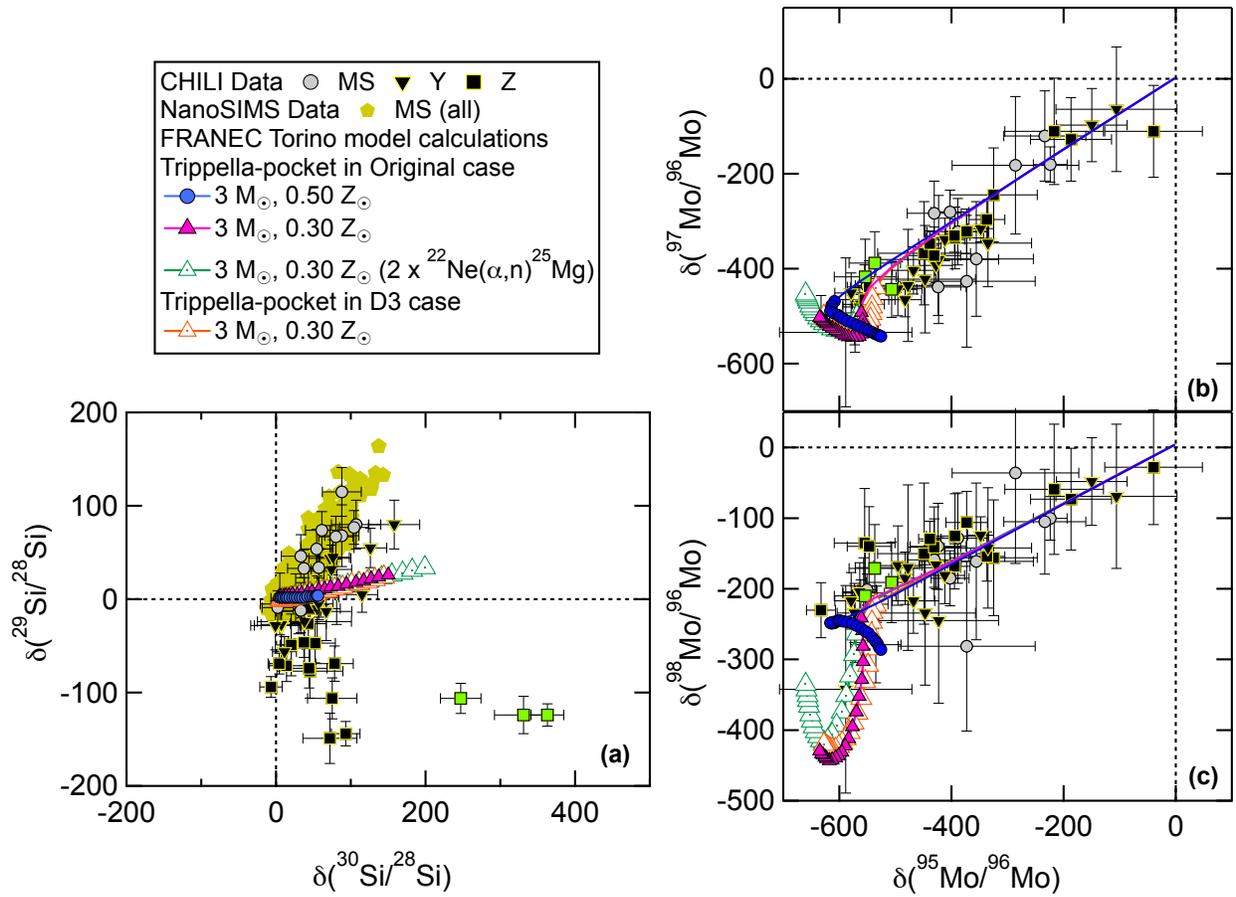

Fig. 7. The same set of plots as in Figs. 4−6 with the grains compared to FRANEC Torino model predictions for 3 $M_\odot$ AGB stars with different metallicities.

The Astrophysical Journal

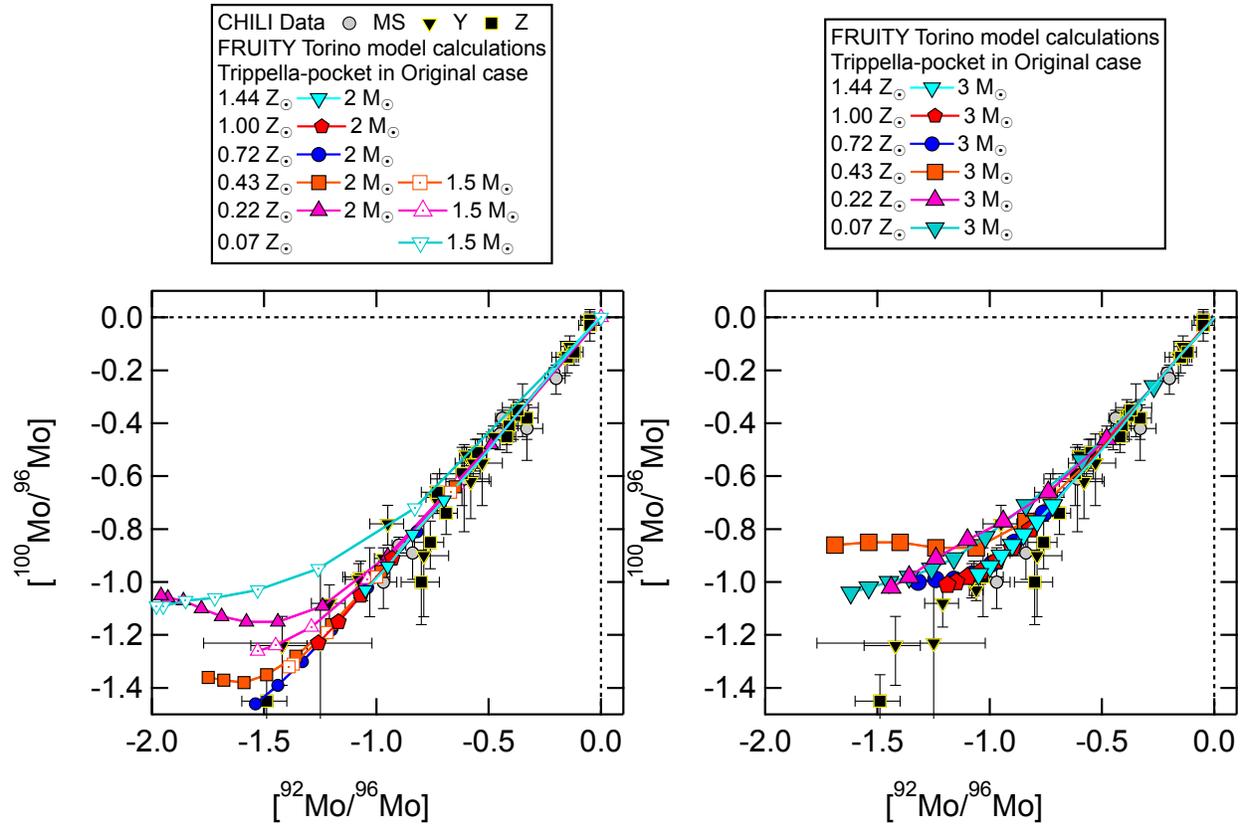

Fig. 8. Three-isotope plot of [$^{100}$Mo/$^{96}$Mo] vs [$^{92}$Mo/$^{96}$Mo] with the grains compared to FRUITY Torino predictions shown in Figs. 2−6.

**Table 1**

Silicon, Niobium, and Molybdenum Neutron Capture Cross Sections, $\sigma_{code}^i = <\sigma^i v> v_{T(30\ keV)} = \sigma_{MACS}^i\ v_T/v_{T(30\ keV)}$ (mbarn)

| Isotope | KADoNiS v0.3 | | | | KADoNiS v1.0 | | | |
|---|---|---|---|---|---|---|---|---|
| | 8 keV | 23 keV | 30 keV | 1 σ err (30 keV) | 8 keV | 23 keV | 30 keV | 1 σ err (30 keV) |
| $^{28}$Si | 0.15 | 1.15 | 1.42 | 9% | 0.23 | 1.10 | 1.42 | 9% |
| $^{29}$Si | 5.56 | 7.39 | 6.58 | 10% | 5.79 | 8.00 | 7.56 | 8% |
| $^{30}$Si | 8.74 | 2.49 | 1.82 | 18% | 4.73 | 2.06 | 1.82 | 18% |
| $^{93}$Nb | 337 | 281 | 266 | 2% | 340 | 292 | 277 | 2% |
| $^{94}$Nb* | 423 | 488 | 482 | 19% | 462 | 419 | 397 | 25% |
| $^{95}$Nb* | 360 | 325 | 310 | 21% | 526 | 479 | 463 | 25% |
| $^{92}$Mo | 98 | 74 | 70 | 14% | 72 | 69 | 67 | 15% |
| $^{93}$Mo*[a] | 336 | 384 | 390 | 25% | 180 | 147 | 139 | 50% |
| $^{94}$Mo | 120 | 104 | 102 | 20% | 114 | 112 | 111 | 14% |
| $^{95}$Mo | 322 | 301 | 292 | 4% | 420 | 388 | 375 | 10% |
| $^{96}$Mo | 150 | 118 | 112 | 7% | 116 | 108 | 105 | 14% |
| $^{97}$Mo | 388 | 353 | 339 | 4% | 421 | 380 | 365 | 15% |
| $^{98}$Mo | 112 | 100 | 99 | 7% | 109 | 93 | 91 | 7% |
| $^{99}$Mo* | 284 | 253 | 240 | 17% | 407 | 379 | 366 | 25% |
| $^{100}$Mo | 111 | 107 | 108 | 13% | 98 | 88 | 86 | 7% |

Note that the KADoNiS v0.3 values were only adopted in the models shown in Fig. 1 and the KADoNiS v1.0 values were adopted in the rest of the models shown in this study.

* denotes unstable isotopes and no experimental measurement available. Given cross sections are derived from theoretical predictions.

[a] The MACS values for $^{93}$Mo are not included in the KADoNiSv0.3 database and are from Rauscher & Thielemann (2000).

**Table 2.** Isotope Data of Type Y and Z Grains.

| Grain | Group | Size (μm²) | $^{12}C/^{13}C$ | $^{14}N/^{15}N$ | $\delta^{29}Si$ (‰) | $\delta^{30}Si$ (‰) | $\delta^{92}Mo$ (‰) | $\delta^{94}Mo$ (‰) | $\delta^{95}Mo$ (‰) | $\delta^{97}Mo$ (‰) | $\delta^{98}Mo$ (‰) | $\delta^{100}Mo$ (‰) |
|---|---|---|---|---|---|---|---|---|---|---|---|---|
| M1-A3-G246 | Y | 1.4×1.1 | 102±2 | 670±115 | −5±33 | 5±26 | −894±9 | −864±12 | −534±19 | −448±26 | −222±25 | −876±11 |
| M1-A3-G346 | Y | 0.7×0.9 | 108±2 | 368±70 | 44±35 | 72±28 | −315±91 | −300±110 | −106±108 | −64±131 | −69±102 | −295±110 |
| M1-A4-G473 | Y | 1.5×1.3 | 119±4 | 2795±797 | −28±37 | 7±36 | −820±15 | −827±19 | −483±29 | −466±35 | −185±36 | −792±21 |
| M1-A5-G1096 | Y | 0.7×1.2 | 117±2 | 394±58 | 8±35 | 81±32 | −888±18 | −850±27 | −496±43 | −443±56 | −167±55 | −832±28 |
| M1-A7-G812 | Y | 0.8×0.9 | 151±4 | 1213±289 | −10±35 | 55±31 | −752±17 | −721±22 | −428±28 | −393±35 | −166±33 | −700±23 |
| M2-A1-G176 | Y | 1.0×1.6 | 112±6 | 397±34 | 45±19 | 76±22 | −939±11 | −898±17 | −573±31 | −518±41 | −234±41 | −917±16 |
| M2-A1-G211 | Y | 0.8×0.6 | 127±7 | 2671±372 | −27±22 | 7±27 | −962±11 | −929±19 | −572±41 | −523±53 | −250±54 | −943±17 |
| M2-A1-G670 | Y | 1.2×1.1 | 112±7 | 633±149 | −28±26 | −1±33 | −708±71 | −765±78 | −423±107 | −380±135 | −244±117 | −716±86 |
| M2-A2-G1140 | Y | 1.2×1.2 | 120±3 | … | 33±22 | 57±24 | −913±6 | −866±9 | −562±14 | −474±19 | −208±19 | −906±7 |
| M2-A2-G154 | Y | 1.2×1.5 | 109±6 | 7602±3867 | −24±47 | 38±63 | −944±39 | −889±70 | −589±118 | −534±156 | −342±146 | −941±56 |
| M2-A2-G234 | Y | 1.2×0.8 | 125±5 | … | 9±21 | 43±23 | −837±47 | −767±71 | −478±94 | −435±118 | −170±117 | −874±51 |
| M2-A2-G262 | Y | 1.2×0.8 | 148±7 | … | 80±26 | 158±34 | −596±57 | −587±70 | −335±77 | −346±91 | −142±84 | −588±70 |
| M2-A2-G652 | Y | 1.3×1.3 | 125±16 | … | −13±24 | 30±30 | −915±12 | −876±18 | −564±30 | −480±40 | −204±40 | −895±16 |
| M3-G1207 | Y | 1.3×1.5 | 122±5 | 451±49 | 55±21 | 126±22 | −592±17 | −581±20 | −348±22 | −315±27 | −124±25 | −568±21 |
| M3-G1250 | Y | 0.8×0.8 | 118±6 | 4070±838 | 32±26 | 74±27 | −733±25 | −706±33 | −468±38 | −403±49 | −217±45 | −720±32 |
| M3-G1261 | Y | 1.3×1.1 | 106±5 | 1784±386 | 5±19 | 115±21 | −666±24 | −652±30 | −412±34 | −339±45 | −181±39 | −648±30 |
| M3-G1386 | Y | 1.3×1.4 | 136±6 | 3233±429 | −5±17 | 57±19 | −269±57 | −291±66 | −150±62 | −97±76 | −48±62 | −232±69 |
| M3-G1712 | Y | 0.9×1.0 | 129±2 | 1027±139 | −56±16 | 11±16 | −916±12 | −867±18 | −579±28 | −451±41 | −217±39 | −897±16 |
| M3-G758 | Y | 0.8×0.9 | 151±7 | 1614±417 | −13±31 | 67±34 | −739±58 | −696±80 | −447±90 | −423±112 | −234±102 | −757±68 |
| M2-A1-G132 | Z | 0.9×0.8 | 56±3 | 2083±335 | −149±27 | 72±36 | −908±22 | −874±33 | −555±53 | −466±73 | −135±77 | −896±30 |
| M2-A1-G301 | Z | 0.9×0.8 | 82±5 | 459±68 | −124±20 | 331±39 | −842±30 | −814±41 | −555±55 | −417±79 | −210±73 | −899±31 |
| M2-A1-G425 | Z | 1.0×1.3 | 83±5 | 374±38 | −69±19 | 78±25 | −968±7 | −924±14 | −633±26 | −495±38 | −230±38 | −965±9 |
| M2-A1-G469 | Z | 0.7×0.7 | 65±4 | 1166±150 | −106±16 | 247±27 | −826±24 | −807±33 | −507±44 | −443±58 | −191±57 | −858±28 |
| M2-A2-G618 | Z | 0.8×1.1 | 48±2 | 1343±491 | −106±22 | 75±33 | −581±29 | −557±36 | −373±37 | −321±47 | −106±43 | −575±35 |
| M2-A2-G650 | Z | 1.2×1.1 | 55±1 | … | −47±21 | 52±27 | −760±49 | −710±70 | −449±83 | −367±109 | −149±100 | −745±63 |
| M2-A2-G654 | Z | 0.9×1.0 | 63±2 | … | −77±29 | 45±44 | −292±83 | −338±94 | −218±87 | −111±112 | −59±91 | −285±99 |
| M2-A2-G782 | Z | 0.5×0.7 | 48±2 | 696±134 | −71±17 | 14±23 | −537±61 | −544±74 | −325±77 | −244±98 | −156±81 | −584±69 |
| M2-A2-G791 | Z | 0.7×0.7 | 46±2 | 2974±612 | −46±16 | 37±21 | −610±39 | −614±47 | −394±51 | −335±65 | −125±60 | −614±47 |
| M2-A2-G908 | Z | 0.8×0.8 | 65±3 | 475±70 | −144±13 | 93±19 | −719±30 | −681±40 | −431±46 | −372±59 | −141±56 | −692±40 |
| M3-G1006 | Z | 0.9×0.7 | 82±1 | 421±48 | −124±12 | 363±22 | −814±27 | −759±38 | −537±45 | −388±65 | −171±61 | −780±36 |
| M3-G1084 | Z | 0.8×0.7 | 51±2 | 867±131 | 12±18 | 64±20 | −244±69 | −249±82 | −187±71 | −127±88 | −73±71 | −254±79 |
| M3-G1672 | Z | 0.9×1.0 | 46±1 | 2273±408 | −69±11 | 4±15 | −797±25 | −758±34 | −547±39 | −438±55 | −140±56 | −820±29 |
| M3-G521 | Z | 1.3×1.3 | 62±3 | 405±84 | −26±22 | 43±24 | −575±24 | −587±28 | −337±31 | −296±39 | −153±34 | −551±30 |
| M3-G561 | Z | 1.2×1.1 | 43±1 | 476±55 | −49±33 | 20±15 | −109±84 | −66±100 | −40±87 | −110±96 | −28±80 | −29±103 |
| M3-G628 | Z | 1.8×2.0 | 40±1 | 2156±466 | −94±11 | −7±15 | −683±14 | −646±18 | −439±19 | −347±25 | −129±24 | −664±17 |
| **M3-G692\*** | **Z** | **0.7×0.6** | **90±4** | **1128±292** | **−10±20** | **45±22** | **−101±102** | **−61±122** | **30±110** | **−29±123** | **45±103** | **−77±120** |
| M3-GB4 | Z | 0.7×0.7 | 75±3 | 1759±291 | −74±21 | 44±24 | −624±20 | −610±25 | −395±27 | −331±34 | −168±31 | −645±24 |

Note: We were not able to obtain $^{14}N/^{15}N$ ratios in six of the 38 grains because of a problem with one of the detectors during one of the NanoSIMS sessions. The delta-notation is defined as $\delta^i A=[(^i A/^j A)_{grain}/(^i A/^j A)_{std}-1]\times 1000$, where A denotes an element, $i$ an isotope of this element, and $j$ the normalization isotope, and $(^i A/^j A)_{grain}$ and $(^i A/^j A)_{std}$ represent the corresponding isotope ratios measured in a sample and the standard, respectively. The normalization isotopes are $^{28}Si$ and $^{96}Mo$. $\delta^i A$ is written as $\delta(^i A/^j A)$ in the figures to further denote the normalization isotope. All values are reported with 2σ errors.

*The correlated Sr, Mo, and Ba isotope ratios of this grain indicate that a large amount of Mo contamination was sampled during the grain analysis.

**Table 3**
Summary of a Complete Model-Data Comparison at Different Metallicities and Masses

| Mass | 0.07 $Z_\odot$ | 0.22 $Z_\odot$ | 0.43 $Z_\odot$ | 0.56 $Z_\odot$ | 0.72 $Z_\odot$ | 1.00 $Z_\odot$ | 1.44 $Z_\odot$ |
|---|---|---|---|---|---|---|---|
| 1.5 $M_\odot$ | ✗ | ✓ | ✓ | ✓ | ✓ | ✓ | ✓ |
| 2.0 $M_\odot$ | ✗ | ✗ | ✗* | ✓ | ✓ | ✓ | ✓ |
| 3.0 $M_\odot$ | ✗ | ✗ | ✗ | ✗ | ✗ | ✓ | ✓ |
| 4.0 $M_\odot$ | … | … | ✗ | O-rich | O-rich | O-rich | O-rich |

**Notes.** Model predictions showing more than 50‰ deviations from the linear trends defined by the grain data in the plot of $\delta^{98}$Mo versus $\delta^{95}$Mo in Figs. 4−6 and are thus inconsistent with the grain data are denoted with a cross mark. Model predictions that show less than 50‰ deviations from the linear trends in Figs. 4−6 and are thus consistent with the grain data are denoted with a check mark. We did not run the two cases for 4.0 $M_\odot$ with lowest metallicities, because $T_{\rm MAX}$ at 0.43 $Z_\odot$ is already too high to match the grain data in Fig. 5. We also did not run FRUITY Torino model calculations for O-rich AGB stars, because SiC is predicted to condense in C-rich environment (Lodders & Fegley 1995).
*This model was excluded based on Fig. 8 instead of the $\delta^{98}$Mo versus $\delta^{95}$Mo plot in Fig. 4.